\documentclass[twocolumn,floats,showpacs,prb,amsfonts,amssymb]{revtex4}

\usepackage{graphicx}
\usepackage{amsmath}        
\usepackage{amsfonts}
\usepackage{amssymb}
\usepackage{bbm}
\usepackage{bm}

\begin{document}

\title{Catching the zitterbewegung}

\author{P. Brusheim}
\email{brusheimp@ihpc.a-star.edu.sg}
\affiliation{Institute of High Performance Computing, 1 Fusionopolis Way, \#16-16 Connexis, Singapore 138632, Singapore.}
\author{H. Q. Xu}
\email{Hongqi.Xu@ftf.lth.se}
\affiliation{Division of Solid State Physics, Lund University, Box 118, S-22100 Lund, Sweden.}

\date{\today}
\begin{abstract}
We demonstrate how the zitterbewegung charge oscillations can be detected through a charge conductance measurement in a three-terminal junction. By tuning the spin-orbit interaction strength or an external magnetic field the zitterbewegung period can be modulated, translating into complementary conductance oscillations in the two outgoing leads of the junction. The proposed experimental setup is within the reach of demonstrated technology and material parameters, and enables the observation of the so far elusive zitterbewegung phenomenon. 
\end{abstract}


\maketitle

\section{Introduction}
The quantum relativistic theory of the electron, explaining the ``peculiar two-valuedness'' or spin angular momentum, was put forward in 1928 by Dirac. While studying free particle solutions to Dirac's theory, Schr\"odinger found that the time dependence of the position operator of an electron traveling through vacuum displays an oscillatory motion around some mean position, named the {\it zitterbewegung}, arising due to an interference between the positive and negative energy states of the Dirac electron. The angular frequency of the oscillation is $\omega=2mc^2/\hbar$ and the amplitude is twice the Compton radius $\lambda = c/\omega = 2r_c = \hbar/mc$, implying an angular momentum $s=mcr_c=\hbar/2$. This equality of the zitterbewegung amplitude and the intrinsic spin angular momentum of the electron has led to the suggestion that indeed the zitterbewegung is the fundamental mechanism behind the intrinsic angular momentum. Hestenes\cite{Hestenes90} extended this idea further by suggesting that the zitterbewegung is a truly fundamental property of the entire quantum reality responsible for the appearance of the complex phase. Establishing whether the zitterbewegung is indeed a real physical observable is therefore of great importance for a deep understanding of the quantum reality. However, due to the extraordinary frequencies and amplitudes set by the electron-positron energy gap, a direct observation of the electron zitterbewegung in vacuum is beyond the reach of standard measurement techniques. 

In the non-relativistic limit the intrinsic angular momentum of the electron is a two-component spinor property. Here, the positive-negative energy interactions of the four-component Dirac equation are folded into spin-dependent terms in the Hamiltonian. For an electron with non-zero momentum, $p$, in an electric field, $E$, this gives rise to a coupling of the spin and spatial degrees of freedom through the spin-orbit interaction (SOI) Hamiltonian $H_{\mathrm{SOI}}=(e\hbar/4m^2c^2)\sigma\cdot E\times p$.\cite{Foldy58} The SOI gives rise to interference properties between the two spin states in a way analogous to the interference between the positive- and negative-energy wave components of the relativistic Dirac particle in vacuum. The prefactor contains the inverse of the electron-positron creation energy ($\approx$ 1 MeV) leading to the difficulty in the observation of effects due to the SOI for a free particle in vacuum. In a semiconducting crystal, however, the relevant energy scale is the conduction-valence band gap which is about six orders of magnitude smaller than the electron-positron gap, making SOI effects of strong influence on the spin dynamics. One thus heuristically expects the zitterbewegung to be of much larger amplitude and lower frequency in crystalline materials.\cite{Zawadzki05,Schliemann06,Winkler07} Here, the electric field which induces the SOI is naturally supplied by the atomic potentials as well as interfacial potentials between different crystalline materials. The SOI derived from the atomic potentials
is called the Dresselhaus SOI and is naturally a fixed material property. The SOI derived from the interfacial field is called the Rashba SOI and can be tuned by, e.g., a voltage applied to a surface gate,\cite{Nitta97,Koga02b,Kohda08} tilting the band structure of a quantum well. It was independently shown in  Refs.~\onlinecite{Schliemann06}, \onlinecite{Brusheim06}, and \onlinecite{Lee05} that zitterbewegung charge oscillations on the scale of 100~nm arise in a semiconducting waveguide with a spin-polarized electron injection in the presence of Rashba SOI. Large spin polarization has been theoretically predicted\cite{Ohe05,Cummings08} and experimentally achieved in three-terminal semiconducting structures.~\cite{Jacob09} The spin separation in these structures rely on the spin-Hall effect induced by the SOI. It was shown in Ref.~\onlinecite{Brusheim06} that the zitterbewegung and the spin Hall effect are intrinsically connected and only differs in injection and detection conditions. It should therefore be possible to measure the zitterbewegung in these structures.

Calculations of the average position of a free (unbound) Dirac particle\cite{Lock79} as well as wave packets in graphene layers and carbon nanotubes\cite{Rusin07} showed that the zitterbewegung oscillations have a transient character with a decay time of $\sim 10^{-12}-10^{-15}$~s. This transient decay occurs due to a spreading of the wave packet, as well as separation of the subpackets corresponding to the positive and negative energy solutions. However, Lock\cite{Lock79} pointed out that in a confined system this spreading is restricted and the transient character would therefore not appear for bound states. An attempt to observe the zitterbewegung should therefore be directed to quantum confined systems where periodic oscillations can be sustained. The approach of studying wave-packet solutions might not be very well suited for the steady-state electron flux in the crystal. Indeed, the electron flux is not a localized ``particle'' but rather a quasi-particle excitation represented by an extended state. A proper analysis of these extended state solutions are thus called for. Previous suggestions for measurements in quantum wires include direct scanning gate\cite{Topinka00,Topinka01,Woodside02,Pioda04} or Kerr rotation measurements\cite{Kato04,Crooker05} of the zitterbewegung oscillations. This would require resolutions of the order of tens of nanometers, which is on the cutting edge of scanning gate techniques and far beyond the capabilities of the Kerr rotation method.
  
\begin{figure}[t]
  \includegraphics[width=8cm]{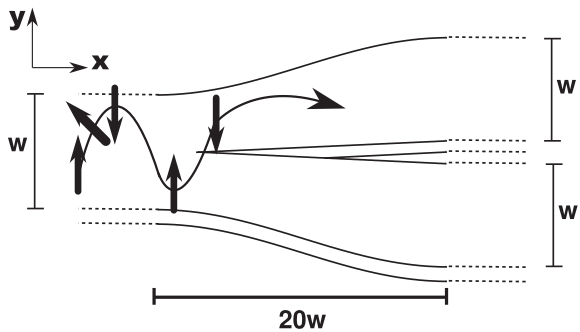}
  \caption{Schematic description of the considered three-terminal junction structure. Arrows indicate schematically the spin-polarization evolution and the quasi-classical zitterbewegung path of electrons.}
\label{str}
\end{figure}

In this paper, we will analytically derive through second order perturbation theory an expression for the zitterbewegung charge oscillation patterns for steady-state quantum transport in a quantum wire. We also perform exact numerical calculations of the transport properties for a three-terminal junction. We will show how a simple charge conductance measurement can be performed to measure the zitterbewegung in such a junction defined in a semiconducting material. The proposed experimental setup, illustrated in Fig.~\ref{str}, employs standard nano-scale lithographic techniques as well as experimentally demonstrated material parameters. 

\section{Zitterbewegung of the electron flux in a quantum wire}

We start by considering a uniform quantum wire (the left part of Fig.~\ref{str}).
The considered quantum wire is defined by a hard-wall confinement potential and is subjected to the Rashba SOI and a magnetic field $B_y$ applied along the $y$ direction. The system is described by the Hamiltonian (with $\hbar=1$)
\begin{eqnarray}
  H &=& \frac{p^2}{2m^*} + V(\bm{r}) + \frac{g^*\mu_B}{2}B_y\sigma_y + \alpha\left(\sigma_xp_y - \sigma_yp_x\right) \nonumber \\
  &=& H_0 + H_S,
\label{Ham}
\end{eqnarray}
where $H_S$ contains all the spin-dependent terms. We will choose the spin quantization axis to be along the magnetic field direction. The scattering state at a given energy in the quantum wire can be written as an expansion of the eigenstates of the wire
\begin{equation}
\Psi = \sum_\alpha a_\alpha \phi_\alpha(y,\sigma) e^{ik_\alpha x}.
\end{equation} 
The analytical solutions for the eigenstates of the spin-orbit coupled quantum wire are not know. We therefore construct the states from second-order perturbation theory where the unperturbed states are the eigenstates of $H_0$:
\begin{equation}
  |n,\sigma^{(0)}\rangle = \sqrt{\frac{2}{w}} \sin\left[\frac{n\pi}{w}(y+1/2)\right]|\sigma\rangle e^{ik x},
\end{equation}
with the eigenenergies
\begin{equation}
  \varepsilon_n(k) = \frac{(n\pi)^2}{w^22m^*} + \frac{k^2}{2m^*}.
\end{equation}
 The unperturbed eigenstates are spin degenerate. However, the total Hamiltonian is symmetric under the $\sigma_yR_y$ operation (where $R_y$ is the reflection operator in the transverse direction). This operation defines a pseudo-spin (generalized parity) which block diagonalizes the Hamiltonian. We can therefore perform the calculations separately in the pseudo-spin spaces $s^+ = \left\{|1\uparrow^{(0)}\rangle, |2\downarrow^{(0)}\rangle, |3\uparrow^{(0)}\rangle, |4\downarrow^{(0)}\rangle,\ldots\right\}$ and $s^- = \left\{|1\downarrow^{(0)}\rangle, |2\uparrow^{(0)}\rangle, |3\downarrow^{(0)}\rangle, |4\uparrow^{(0)}\rangle,\ldots\right\}$.  

The general form of the first-order correction to the $s^\pm$ pseudo-spin states read
\begin{equation}
  |n\gamma\in s^{\pm(1)}\rangle = \sum_{\{m\sigma\neq n\gamma\} \in s^\pm} | m\sigma^{(0)}\rangle \frac{c_{m\sigma,n\gamma}}{\varepsilon_n-\varepsilon_m}, 
\end{equation}
where we have introduced the shorthand notation $c_{i\sigma j\sigma'} = \langle i,\sigma^{(0)}|\alpha\sigma_xp_y + \sigma_y(g^*\mu_B B_y/2 - \alpha p_x)|j,\sigma'^{(0)}\rangle \in \mathbbm{R}$. The general form of the second-order correction to the $s^\pm$ pseudo-spin states read
\begin{eqnarray}
   |n\gamma\in s^{\pm(2)}\rangle &=& \sum_{\substack{\{m\sigma\neq n\gamma\} \in s^\pm \\ \{l\sigma'\neq n\gamma\} \in s^\pm}} | m\sigma^{(0)}\rangle \frac{c_{m\sigma,l\sigma'}c_{l\sigma',n\gamma}}{(\varepsilon_n-\varepsilon_m)(\varepsilon_n-\varepsilon_l)} \nonumber \\
   &-& \sum_{\{m\sigma\neq n\gamma\} \in s^\pm} | m\sigma^{(0)}\rangle \frac{c_{n\gamma,n\gamma}c_{m\sigma,n\gamma}}{(\varepsilon_n-\varepsilon_m)^2} \nonumber \\
   &-& \frac{1}{2}\sum_{\{m\sigma\neq n\gamma\} \in s^\pm} | n\gamma^{(0)}\rangle\frac{c_{n\gamma,m\sigma}c_{m\sigma,n\gamma}}{(\varepsilon_n-\varepsilon_m)^2}.  
\end{eqnarray} 
Writing out explicitly the first-order correction to the lowest $s^+$ pseudo-spin state we have
\begin{eqnarray}
  |1s^{+(1)}\rangle &=& \frac{c_{2\downarrow 1\uparrow}}{\varepsilon_{1}-\varepsilon_{2}}|2\downarrow^{(0)}\rangle + \frac{c_{3\uparrow 1\uparrow}}{\varepsilon_{1}-\varepsilon_{3}}|3\uparrow^{(0)}\rangle \nonumber \\
  &+& \frac{c_{4\downarrow 1\uparrow}}{\varepsilon_{1}-\varepsilon_{4}}|4\downarrow^{(0)}\rangle + \ldots \nonumber \\
  &\approx& \frac{\alpha}{wE_1}\left[\frac{8}{9}|2\downarrow^{(0)}\rangle + \frac{16}{225} |4\downarrow^{(0)}\rangle\right],
\label{firstorder}
\end{eqnarray}
where we have truncated the converging series and set $E_1=\pi^2/2m^*w^2$. The second term in Eq.~(\ref{firstorder}), $c_{3\uparrow 1\uparrow}$, vanishes due to the orthogonality of the unperturbed states. Similarly, for the lowest $s^-$ pseudo-spin state, the first order correction reads
\begin{eqnarray}
|1s^{-(1)}\rangle &\approx& \frac{c_{2\uparrow 1\downarrow}}{\varepsilon_{1}-\varepsilon_{2}}|2\uparrow^{(0)}\rangle + \frac{c_{4\uparrow 1\downarrow}}{\varepsilon_{1}-\varepsilon_{4}} |4\uparrow^{(0)}\rangle \nonumber \\
&=& -\frac{c_{2\downarrow 1\uparrow}}{\varepsilon_{1}-\varepsilon_{2}}|2\uparrow^{(0)}\rangle - \frac{c_{4\downarrow 1\uparrow}}{\varepsilon_{1}-\varepsilon_{4}} |4\uparrow^{(0)}\rangle,
\end{eqnarray}
where in the last equality we have used the fact that $c_{i\sigma j\sigma'} = -c_{i\bar{\sigma}j\bar{\sigma'}}$, with the bar indicating a spin flip. We can then put both first order corrections to the two pseudo-spin states on a short form as
\begin{eqnarray}
   |1s^{+(1)}\rangle &\approx& c_2^{(1)}|2\downarrow^{(0)}\rangle + c_4^{(1)}|4\downarrow^{(0)}\rangle, \nonumber \\
   |1s^{-(1)}\rangle &\approx& -c_2^{(1)}|2\uparrow^{(0)}\rangle - c_4^{(1)}|4\uparrow^{(0)}\rangle.
   \label{symmFirstOrd}
\end{eqnarray}
The second order correction to the lowest $s^+$ pseudo-spin state reads
\begin{eqnarray}
  &&|1s^{+(2)}\rangle \approx \left[ \frac{(c_{2\downarrow 2\downarrow}-c_{1\uparrow 1\uparrow})c_{2\downarrow 1\uparrow}}{(\varepsilon_1-\varepsilon_2)^2}
\right] |2\downarrow^{(0)}\rangle \nonumber \\
  &+& \left[ \frac{c_{3\uparrow 2\downarrow}c_{2\downarrow 1\uparrow}}{(\varepsilon_1-\varepsilon_3)(\varepsilon_1-\varepsilon_2)}+\frac{c_{3\uparrow 4\downarrow}c_{4\downarrow 1\uparrow}}{(\varepsilon_1-\varepsilon_3)(\varepsilon_1-\varepsilon_4)} \right] |3\uparrow^{(0)}\rangle \nonumber \\
  &+& \left[ 
\frac{(c_{4\downarrow 4\downarrow}-c_{1\uparrow 1\uparrow})c_{4\downarrow 1\uparrow}}{(\varepsilon_1-\varepsilon_4)^2} \right] |4\downarrow^{(0)}\rangle . \nonumber \\
\end{eqnarray}
A similar expression can also be derived for the second order correction to the lowest $s^-$ pseudo-spin state.
Here, due to the second order nature, the coefficients are invariant under the spin flip and we can thus put the two 
second-order corrections on the short form as
\begin{eqnarray}
  |1s^{+(2)}\rangle &\approx& c_2^{(2)}|2\downarrow^{(0)}\rangle + c_3^{(2)}|3\uparrow^{(0)}\rangle + c_4^{(2)}|4\downarrow^{(0)}\rangle, \nonumber \\
  |1s^{-(2)}\rangle &\approx& c_2^{(2)}|2\uparrow^{(0)}\rangle + c_3^{(2)}|3\downarrow^{(0)}\rangle + c_4^{(2)}|4\uparrow^{(0)}\rangle. \nonumber \\
\label{symmSecondOrd}
\end{eqnarray} 
We want to study the zitterbewegung pattern for transport in the two lowest eigenchannels of the wire and, therefore, calculate the expectation value of the transverse coordinate,
\begin{eqnarray}
  \langle y \rangle(x) &=& \sum_{\alpha\beta} a_\beta^*a_\alpha \langle\beta|y|\alpha\rangle e^{i(k_\alpha-k_\beta)x} \nonumber \\
  &=& |a_{1s^+}|^2\langle 1s^+|y|1s^+\rangle + |a_{1s^-}|^2\langle 1s^-|y|1s^-\rangle \nonumber \\
  &+& 2Re\left[a_{1s^-}^*a_{1s^+}\langle 1s^-|y|1s^+\rangle e^{i(k_{1s^+}-k_{1s^-})x}\right]. \nonumber \\
\label{y}
\end{eqnarray}
But since the spatial coordinate operator $y$ is of odd parity we have $\langle 1,s^+|y|1,s^+\rangle = \langle 1,s^-|y|1,s^-\rangle=0$. Thus, Eq.~(\ref{y}) implies that there is no uniform displacement of the wave function in the $y$-direction derived from the spin-orbit interaction or the in-plane magnetic field $B_y$ (this is true to all orders).
We are hence left with calculating the term
\begin{eqnarray}
  \langle 1s^-|y|1s^+\rangle = \langle 1\downarrow^{(0)}|y|1\uparrow^{(0)}\rangle &+&  \langle 1\downarrow^{(0)}|y|1s^{+(1)}\rangle \nonumber \\
  + \langle 1s^{-(1)}|y|1\uparrow^{(0)}\rangle &+& \langle 1s^{-(1)}|y|1s^{+(1)}\rangle \nonumber \\
  +  \langle 1\downarrow^{(0)}|y|1s^{+(2)}\rangle &+& \langle 1s^{-(2)}|y|1\uparrow^{(0)}\rangle \nonumber \\
  + \langle 1s^{-(2)}|y|1s^{+(1)}\rangle &+& \langle 1s^{-(1)}|y|1s^{+(2)}\rangle \nonumber \\
  + \langle 1s^{-(2)}|y|1s^{+(2)}\rangle.&&
\label{terms}
\end{eqnarray}
The first term is obviously zero. From Eq.~(\ref{symmFirstOrd}) we can easily see that the second and third terms are mutually cancelling. Furthermore, from Eq.~(\ref{symmFirstOrd}) we see that the first order corrections to the lowest $s^+$ states only contains unperturbed spin-down components and vice versa for the corrections to the lowest $s^-$ states. This means that the fourth term is identically zero. From Eq.~(\ref{symmSecondOrd}) we see that the fifth and sixth terms in Eq.~(\ref{terms}) combine to a second-order non-zero contribution. The seventh, eighth, and ninth terms are third- and fourth-order contributions and will hence be discarded. We are then left with the result
\begin{eqnarray}
   &&\langle 1s^-|y|1s^+\rangle \nonumber \\
   &=& 2c_2^{(2)}\langle 1\downarrow^{(0)}|y|2\downarrow^{(0)}\rangle + 2c_4^{(2)}\langle 1\downarrow^{(0)}|y|4\downarrow^{(0)}\rangle \nonumber \\
   &=& \frac{2^{8}\alpha}{3^5\pi^2E_1^2}\left[2\alpha k - g^*\mu_BB_y \right] \nonumber \\
  &+& \frac{2^{10}\alpha}{15^5\pi^2E_1^2}\left[2\alpha k - g^*\mu_BB_y\right].
\end{eqnarray}
The second term can safely be neglected since it represents a correction of the order $c_4^{(2)}/c_2^{(2)}=4/5^5$. 

The coefficients $a_\alpha$ in Eq.~(\ref{y}) are determined through the injection boundary condition. We consider the condition that electrons are injected into the wire from a non-interacting semi-infinite lead. We further assume that the spin-orbit interaction is adiabatically increased towards the finite value in the wire. If the injected electrons are polarized in the out-of plane $z$ direction we have $a^z_{1s^+} = a^z_{1s^-} = 1/\sqrt{2}$. Similarly, for electrons injected with a polarization in the longitudinal $x$ direction we have $a^x_{1s^+} = 1/\sqrt{2}$ and $a^x_{1s^-} = i/\sqrt{2}$ and for electrons polarized in the transverse $y$ direction we have $a^y_{1s^+} = 1$ and $a^y_{1s^-} = 0$. This gives the zitterbewegung pattern
\begin{eqnarray}
  \langle y \rangle^z(x) &=& \frac{2^{8}\alpha}{3^5\pi^2E_1^2}\left[2\alpha k - g^*\mu_BB_y \right] \nonumber \\
  &\times& \cos[(k_{1s^+}-k_{1s^-})x]. \nonumber \\
  \langle y \rangle^x(x) &=& \frac{2^{8}\alpha}{3^5\pi^2E_1^2}\left[2\alpha k - g^*\mu_BB_y\right] \nonumber \\
  &\times& \sin[(k_{1s^+}-k_{1s^-})x]. \nonumber \\
  \langle y \rangle^y(x) &=& 0.
  \label{zitterB}
\end{eqnarray}
Our derivation of Eq.~(\ref{zitterB}) has shown that the steady-state zitterbewegung pattern for the electron flux in a quantum wire is a second order process in the spin-orbit interaction. It arises through interference between the spin-split eigenstates at a finite $k$ vector. The amplitude of the oscillations depends on the SOI strength, the magnetic field applied, and the quantization energy of the wire. Furthermore, it is periodic with a period proportional to the inverse of the $k$ vector difference between the split states, $1/\Delta k$. In order to find this period we calculate the eigenenergy corrections to second order
\begin{equation}
  E_{1s^+} \approx \frac{\pi^2}{2m^*w^2} + \frac{k^2}{2m^*} + \frac{g^*\mu_BB_y}{2} - \alpha k - 5\frac{\alpha^2 m^*}{\pi^2},
\end{equation}
which gives us
\begin{eqnarray}
  && k_{1s^+} \approx \alpha m^* \nonumber \\
  && \pm\sqrt{2(\alpha m^*)^2 + 2m^*E_{1s^+}- \frac{\pi^2}{w^2} - m^*g^*\mu_BB_y}. \;\;
\end{eqnarray}
Similarly, for the lowest $s^-$ pseudo-spin state, we have
\begin{eqnarray}
  && k_{1s^-} \approx -\alpha m^* \nonumber \\
  &&\pm\sqrt{2(\alpha m^*)^2 + 2m^*E_{1s^-}- \frac{\pi^2}{w^2} + m^*g^*\mu_BB_y}. \;\;
\end{eqnarray}
The linear response transport is taking place at the Fermi energy $E_F$. Setting both pseudo-spin energies to $E_F$ and considering the positive branch we have the wave-vector difference
\begin{eqnarray}
  && k_{1s^+} - k_{1s^-} = \Delta k \approx 2\alpha m^* \nonumber \\
  && + \sqrt{2(\alpha m^*)^2 + 2m^*E_F - \frac{\pi^2}{w^2} - m^*g^*\mu_BB_y} \nonumber \\
  && - \sqrt{2(\alpha m^*)^2 + 2m^*E_F - \frac{\pi^2}{w^2} + m^*g^*\mu_BB_y}.\;\;
  \label{freq}
\end{eqnarray}
In the limit of $2m^*E_F - \frac{\pi^2}{w^2}\gg |2(\alpha m^*)^2\pm m^*g^*\mu_BB_y|$ we can expand the square roots to first order 
giving 
\begin{equation}
  \Delta k \approx m^*\left(2\alpha - \frac{g^*\mu_BB_y}{\sqrt{ 2m^*E_F - \pi^2/w^2}}\right).
\label{freqExpand}
\end{equation}
The zitterbewegung period can hence be tuned by changing the SOI strength as well as the external in-plane magnetic field. The magnetic field alone causes a rotation of the spin angular momentum (for $x$ or $z$ polarized injections) but does not generate a spatial charge oscillation. When the spin and orbital degrees are coupled through the SOI, the rotation of the spin angular momentum couples into a spatial charge oscillation. Quantum mechanically this manifests itself in interference between the spin-split states at the Fermi energy. The degree of splitting is tuned by the SOI strength and the magnetic field, creating a modulation in the oscillation period.

\section{Zitterbewegung in a three-terminal junction: Exact numerical solutions}
In order to devise an experiment to measure this effect we consider the three-terminal junction structure, illustrated in Fig.~\ref{str}, to be defined in a two-dimensional electron gas formed inside a semiconductor heterostructure. Manufacturing of such quantum structures can routinely be made by, e.g., state of the art lithographic techniques.\cite{Jacob09,Shorubalko01,Worschech01,Fuhrer06} On top of the structure, a metal gate is placed, enabling tunability of the electric field in the heterostructure and, hence, the Rashba SOI strength. Tunability of the Rashba SOI strength in the ranges of 6-9~meV~nm and of 1.5-5~meV~nm has been demonstrated in InGaAs quantum wells.\cite{Nitta97,Kohda08} The junction is injected with spin-polarized electrons from a spin-filter device.\cite{Ohe05,Cummings08,Jacob09,Koga02,Hickey05,Hickey08,Zhai06,Brusheim06b} The waveguide presents a quantum confined structure in the transverse direction eliminating the detrimental effects of zitterbewegung transients expected in extended systems.\cite{Lock79,Rusin07} The results of Refs.~\onlinecite{Schliemann06}, \onlinecite{Brusheim06} and \onlinecite{Lee05} as well as Eq.~(\ref{zitterB}) show that the injected carriers will exhibit a transverse oscillatory charge pattern in the waveguide. Intuitively, depending on the period of oscillations the carriers will then be captured with unequal probabilities in the upper and lower branches of the junction, as indicated in Fig.~\ref{str}. Now, according to Eq.~(\ref{freqExpand}) the period of oscillations can be tuned through $\alpha$ or $B_y$ and we therefore expect a periodic modulation of scattering into the upper and lower branches, respectively, as we tune the SOI strength or an external magnetic field. By measuring the conductances between the central and the upper branches and between the central and lower branches, respectively, we should be able to detect the zitterbewegung oscillations as a complimentary oscillating conductance signal. To study this we calculate the  spin-resolved quantum transport properties in such a three-terminal structure by adopting an exact numerical scattering-matrix method developed in Refs.~\onlinecite{Xu94-95,Csontos02-03,Zhang05,Brusheim08}.

\begin{figure}[t]
  \includegraphics[width=8cm]{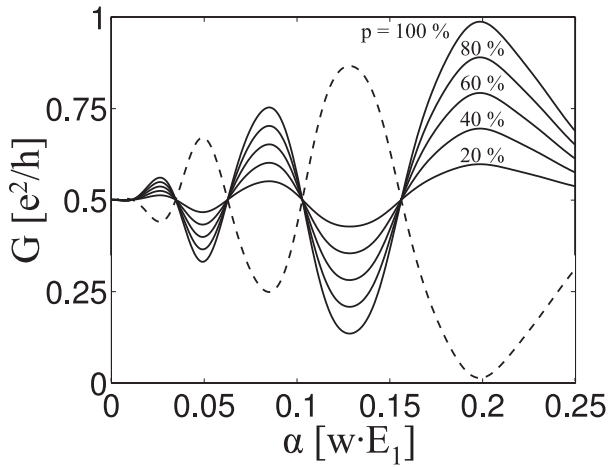}
  \caption{Charge conductance at $B_y=0$ into the upper (dashed line) and lower (solid line) leads as a function of the SOI strength, for carriers injected into the central lead with different degrees of spin polarization along the $z$ direction. The zitterbewegung oscillations are detected as two out-of-phase conductance oscillations. Spin-orbit interaction strength is given in terms of the width $w$ and the lowest transverse eigenenergy $E_1=\pi^2/2m^*w^2$ of the leads. The Fermi energy is set at $E_F=3$~$E_1$.}
  \label{G}
\end{figure}

The SOI is non-zero and spatially uniform in the central region of the structure shown in Fig.~\ref{str}. The perfect leads (not shown in Fig.~\ref{str}) attached to the branches are assumed to have vanishing SOI. Since we shall consider charge transport, this assumption will only induce slight scattering at the lead-branch interface and will hence not change the physics we are studying here. 
Figure~\ref{G} shows the numerically calculated charge conductance into the two outgoing leads (solid and dashed lines, respectively) of the three-terminal structure at $B_y=0$ as a function of the SOI strength. 
The different solid lines represents varying degree of polarization of the injected carriers ranging from a full spin polarization $p=100$\% to a weak spin polarization $p=20$\%.   
As the SOI strength is modulated the charge conductance displays an oscillating behavior. The oscillations of the charge conductances into the upper and lower leads are exactly out of phase and increase in amplitude with increasing the SOI strength. Going from $\alpha\approx 0.13$~$wE_1$ to $\alpha\approx 0.2$~$wE_1$ half a period of oscillation can be achieved. For an InGaAs waveguide of width $w=150$~nm, this corresponds to 6.5~meV~nm~$\leq \alpha \leq$~10 meV~nm, translating to a gate voltage sweep of $\Delta V_g\approx 7$~V (Ref. \onlinecite{Kohda08}), within the reach of the tunability demonstrated in Ref.~\onlinecite{Nitta97}. 

\begin{figure}[t]
  \includegraphics[width=8cm]{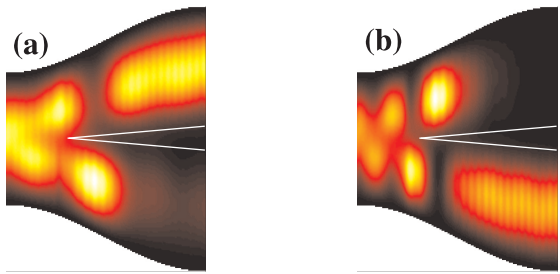}
  \caption{Charge distributions for an electron injection with an initial spin-polarization along the $z$-direction in the three-terminal junction with the SOI strength of (a) $\alpha = 0.129$~$wE_1$ and (b) $\alpha = 0.199$~$wE_1$. The change in the SOI strength results in a modulation in the zitterbewegung period, leading to a capture of electrons in the upper or lower branch.}
  \label{W}
\end{figure}

To demonstrate that these alternating conductance oscillations arise due to the modulation of the zitterbewegung period, we plot in Fig.~\ref{W} the charge probability distribution of the electron wave in the three-terminal junction structure. 
In Fig.~\ref{W}(a) the calculation is made for the SOI strength $\alpha=0.129$~$wE_1$, i.e., at an extreme point of the conductance to the upper lead. As the initially spin-polarized electrons travel along the central branch the SOI induces zitterbewegung oscillations in the charge distribution. When the electrons reach the junction, the central divider captures the oscillations into the upper branch,  resulting in a high conductance to that lead. Conversely, in Fig.~\ref{W}(b) the calculation is made for the SOI strength $\alpha=0.199$~$wE_1$, i.e., at the half-period shift of the extreme point of the conductance to the upper lead. Here the zitterbewegung charge oscillation period is increased and the central divider captures the oscillations into the lower branch. 

\begin{figure}[t]
  \includegraphics[width=8cm]{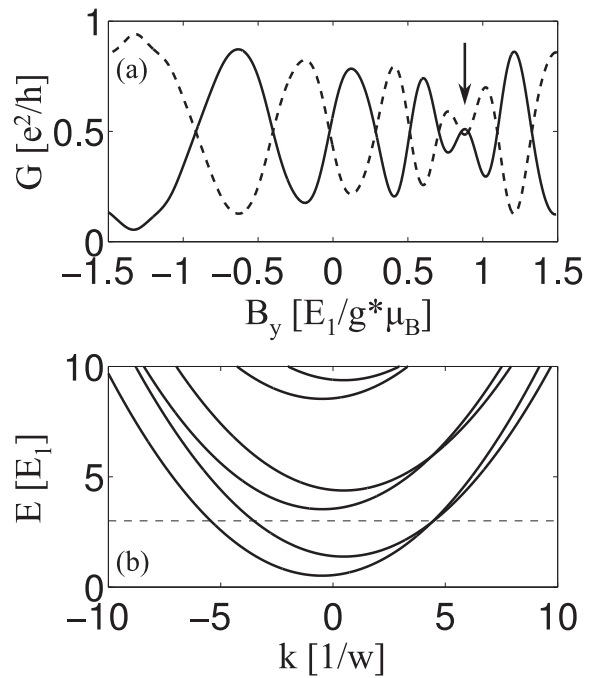}
  \caption{(a) Charge conductance into the upper (dashed line) and lower (solid line) leads as a function of the magnetic field, for carriers with initial spin-polarization along the $z$-direction injected into the central lead at the Fermi energy $E_F=3$~$E_1$. The zitterbewegung oscillations are detected as two out-of-phase conductance oscillations. At $B_y\approx 0.88$~$E_1/g^*\mu_B$ (marked by arrow) the SOI and Zeeman fields are mutually cancelling, creating equal conductances. (b) Band structure of a uniform wire for $B = 0.88$~$g^*\mu_B/E_1$. At the Fermi energy $E_F=3$~$E_1$ (dashed line), the $k$ vector difference of the two lowest forward-propagating states is $\Delta k = 0$ due to the cancelling SOI and Zeeman fields. In both panels $\alpha = 0.1$~$wE_1$.}
  \label{GB}
\end{figure}

The zitterbewegung period can also be tuned by the in-plane magnetic field. In Fig.~\ref{GB}(a) we show the calculated conductances into the two outgoing branches as a function of the magnetic field at the fixed SOI strength $\alpha=0.1$~$wE_1$. At positive fields the conductance show regular oscillations with changing $B_y$. 
At $B_y\approx 0.88$~$E_1/g^*\mu_B$ (marked by an arrow) we can see that the conductance into both outgoing branches are equal (two extreme points). We can understand this from Eq.~(\ref{zitterB}) which says that the zitterbewegung amplitude is proportional to the vector addition of the SOI and Zeeman fields. At the point $2\alpha k = g^*\mu_BB_y$ the two fields are mutually cancelling and the zitterbewegung patterns disappear. Taking the $k$ vector to be the unperturbed value at the Fermi energy, this occurs at $B_y = 2\alpha\pi\sqrt{2}/wg^*\mu_B \approx 0.89$~$E_1/g^*\mu_B$. This explains the equal conductance into both outgoing branches. Since the fields are cancelling we also expect the zitterbewegung period, given by $2\pi/\Delta k$, to be infinite at this point. By inspecting the band structure for a quantum wire [found by exact numerical diagonalization of Eq.~(\ref{Ham})] shown in Fig.~\ref{GB}(b) we can see that at the Fermi energy (marked by a dashed line) the two forward propagating states with positive $k$ number have $\Delta k=0$ at $B_y \approx 0.88$~$E_1/g^*\mu_B$, which indeed implies an infinite zitterbewegung period. The suppression of the zitterbewegung oscillations at this point is hence related to the mutual cancellation of the fields and the subsequent restoration of the time-reversal symmetry and recovery of spin as a good quantum number in the system.\cite{Brusheim07} In the negative magnetic field domain the SOI and Zeeman fields are mutually constructive and there is hence no such degeneracy point.   

\begin{figure}[t]
  \includegraphics[width=8cm]{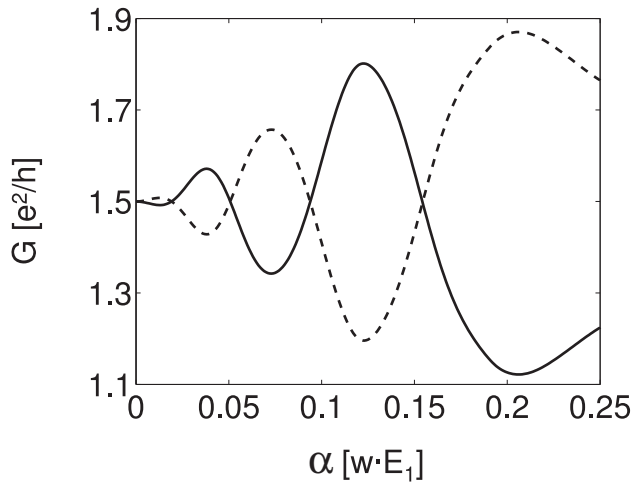}
  \caption{Same as Fig.~\ref{G} except that the Fermi energy is set at $E_F=12$~$E_1$, corresponding to a multi-channel electron transport through the three-terminal junction structure.}
  \label{GE12}
\end{figure}

The above calculations were in the single-channel transport regime. For a multi-channel transport the zitterbewegung patterns becomes increasingly complicated.\cite{Brusheim06} However, from the calculated conductances in the multi-channel transport regime (Fig.~\ref{GE12}), we see that the tell-tale complementary oscillations akin to the zitterbewegung can nevertheless still be clearly discerned.       

\section{Conclusion}
We analytically derived the zitterbewegung charge patterns for the steady-state electron flux in a quantum wire. The zitterbewegung is a second order process in the spin-orbit interaction and its spatial period and amplitude can be shifted by tuning the spin-orbit interaction strength or an external magnetic field. This was confirmed by exact numerical calculations of the charge distribution and the conductance. By performing a charge conductance measurement in a three-terminal semiconductor nanostructure it is possible to detect the zitterbewegung charge oscillations, that has eluded researchers for almost 80 years, as a complimentary conductance oscillation in the outgoing arms. This oscillation should be detectable even for injected carriers with spin-polarization as low as 20\%. Due to the relative simplicity of the set up and the readily available enabling technologies we expect that such a measurement could be performed immediately. Our proposal has a great advantage to, e.g., a two-dimensional electron system defined in a quantum well, which suffers from time decay of the zitterbewegung, and can sustain periodic oscillations due to the quantum confinement. By performing the suggested experiment the question regarding the reality of the zitterbewegung could be settled, and thus bring new light on the very underlying nature of quantum reality.

\section{Acknowledgments}
P. B. acknowledges useful discussions with Jan Jacob. H. Q. X. acknowledges the Swedish
Research Council (VR) for financial support.

\end{document}